\documentclass[review]{elsarticle}

\usepackage{lineno,hyperref}
\modulolinenumbers[5]

\journal{Acta Astronautica}

\usepackage{graphicx} 
\usepackage{subfigure}
\usepackage{booktabs} 
\usepackage{array} 
\usepackage{paralist} 
\usepackage{amsmath} 
\usepackage{amsthm}
\usepackage{mdwlist} 
\usepackage{setspace}
\usepackage{natbib}
\usepackage{epstopdf}

\begin{document}

\begin{frontmatter}

\title{The optimisation of low-acceleration interstellar relativistic rocket trajectories using genetic algorithms}

\author{Kenneth K H Fung\,\tnoteref{mytitlenote}$^{,\mathrm{a,b}}$, Geraint F Lewis\,$^\mathrm{a}$, Xiaofeng Wu\,$^\mathrm{b}$}
\tnotetext[mytitlenote]{Corresponding author.}
\ead{kfun2342@uni.sydney.edu.au, geraint.lewis@sydney.edu.au, xiaofeng.wu@sydney.edu.au}

\address{$^\mathrm{a}$\,Sydney Institute for Astronomy, School of Physics, A28, The University of Sydney, NSW 2006, Australia}
\address{$^\mathrm{b}$\,School of Aerospace, Mechanical and Mechatronic Engineering, J07, The University of Sydney, NSW 2006, Australia}

\begin{abstract}
A vast wealth of literature exists on the topic of rocket trajectory optimisation, particularly in the area of interplanetary trajectories due to its relevance today. Studies on optimising interstellar and intergalactic trajectories are usually performed in flat spacetime using an analytical approach, with very little focus on optimising interstellar trajectories in a general relativistic framework. This paper examines the use of low-acceleration rockets to reach galactic destinations in the least possible time, with a genetic algorithm being employed for the optimisation process. The fuel required for each journey was calculated for various types of propulsion systems to determine the viability of low-acceleration rockets to colonise the Milky Way. The results showed that to limit the amount of fuel carried on board, an antimatter propulsion system would likely be the minimum technological requirement to reach star systems tens of thousands of light years away. However, using a low-acceleration rocket would require several hundreds of thousands of years to reach these star systems, with minimal time dilation effects since maximum velocities only reached about $0.2c$. Such transit times are clearly impractical, and thus, any kind of colonisation using low acceleration rockets would be difficult. High accelerations, on the order of 1\,g, are likely required to complete interstellar journeys within a reasonable time frame, though they may require prohibitively large amounts of fuel. So for now, it appears that humanity's ultimate goal of a galactic empire may only be possible at significantly higher accelerations, though the propulsion technology requirement for a journey that uses realistic amounts of fuel remains to be determined.

\end{abstract}

\begin{keyword}
\texttt{Interstellar trajectory optimization, general relativity, genetic algorithm, Milky Way}
\end{keyword}

\end{frontmatter}

\linenumbers

\section{Introduction}

A wealth of literature exists on optimising space trajectories, in particular interplanetary trajectories due to its application in the near-future. A majority of the research focuses on optimising trajectories for a specific propulsion system, rather than for a general propulsion system that utilises the rocket equation. Solar sails appear to be the favourite propulsion candidate for trajectory optimisation due to the fact that there is no fuel consumption, hence considerably simplifying the analysis: \cite{Cassenti1997} used basic calculus to optimise the solar system exit speed for a spacecraft using a solar sail; \cite{Zeng2011} optimised interplanetary solar sail trajectories with respect to the flight time using particle swarm optimisation; \cite{Dachwald2004,Dachwald2005} used evolutionary neurocontrol to optimise low-thrust interplanetary trajectories; \cite{Kluever1996} used sequential quadratic programming to optimise the flight time for a small spacecraft to reach the edge of the heliosphere using solar and nuclear electric propulsion systems; and \cite{Abdelkhalik2007} used a genetic algorithm to optimise the fuel consumption during orbital transfers. However, solar sails are not practical for interstellar travel since they require a constant external source of energy, which is not always present in the expanse of interstellar space. Research conducted in optimising interstellar trajectories have mostly been performed within a Newtonian model, thereby simplifying the analysis by ignoring the relativistic effects of time dilation. 

The discovery that time is relative has raised many interesting discussions, and has produced a plethora of literature on its effect on interstellar travel. Within the scientific community, many authors have examined the effects of time dilation whilst travelling interstellar and intergalactic distances, though all but a few of the calculations were performed in flat spacetime. \cite{Heyl2005}, \cite{Rindler1960}, and \cite{Kwan2010} considered the effect of an expanding universe when traversing intergalactic distances, and showed that a constant acceleration is necessary if one wishes to reach nearby galaxies within human lifetimes (though this is sensitive to the cosmological parameters used). In the currently favoured cosmological concordance model, a rocketeer accelerating at a constant rate of $g=9.81\,$m\,s$^{-2}$ is able to reach 99\% of the way to the edge of the universe well within a human lifetime \citep{Kwan2010}, though upon return, many billions of years would have passed for those living on Earth. 

Optimising an interstellar trajectory is an extremely complex and difficult task, and producing the correct solution may not always be possible. Almost all attempts consider either a Newtonian or special relativistic approach, as a general relativistic approach compounds the difficulty of the task. \cite{Henriques2012} derived the optimality conditions for rocket trajectories in general relativity, though it was done from an analytical approach and did not consider any specific trajectories. To date, very little research has been performed on optimising interstellar trajectories in a general relativistic framework.

\section{Theory}
\subsection{General Theory of Relativity}

The assumption of the constancy of the speed of light $c$ means that space and time could be unified into a single coordinate system where the position of a particle is $x^\alpha=(t,x,y,z)$. In the curved spacetimes of general relativity, a rocketeer will experience kinematic and gravitational time dilation. The proper time $\tau$ they experience is dependent on their velocity magnitude as well as the local spacetime geometry, described by a metric tensor $g_{\alpha\beta}$.

The equations of motion of a traveller in curved spacetime in Einstein summation convention is given by
\begin{gather}
\frac{d^2x^\alpha}{d\tau^2}=-\Gamma_{\beta\gamma}^\alpha\frac{dx^\beta}{d\tau}\frac{dx^\gamma}{d\tau} + a^\alpha
\label{geoaccel}
\end{gather}
where $a^\alpha=(a^t,a^x,a^y,a^z)$ is the four-acceleration (the relativistic analogue of three-acceleration), and the Christoffel symbols $\Gamma_{\beta\gamma}^\alpha$ describes the local spacetime geometry \cite{Hartle2003},
\begin{gather}
g_{\alpha\delta}\Gamma_{\beta\gamma}^\delta=\frac{1}{2}\left(\frac{\partial g_{\alpha\beta}}{\partial x^\gamma}+\frac{\partial g_{\alpha\gamma}}{\partial x^\beta}-\frac{\partial g_{\beta\gamma}}{\partial x^\alpha}\right)
\label{christeqn}
\end{gather}
The parameters of a particle are related through two normalisation conditions:
\begin{gather}
g_{\alpha\beta}u^\alpha u^\beta=-c^2\label{norm1}\\
g_{\alpha\beta}u^\alpha a^\beta=0\label{norm2}
\end{gather}
where $u^\alpha=(u^t,u^x,u^y,u^z)$ is the four-velocity (the relativistic analogue of Newtonian three-velocity).

\subsection{Milky Way Mass Model}
To model the effect of spacetime curvature due to the mass of the Milky Way, the static weak field metric will be used, which describes the spacetime geometry in a weak, time-independent, gravitational field, such as that of the Milky Way \citep{Gasperini2013}. The static weak field depends on the Newtonian gravitational potential $\Phi$, and is described by the metric
\begin{gather}
g_{\alpha\beta}=\mathrm{diag}\left(-c^2\left(1+\dfrac{2\Phi}{c^2}\right),~1-\dfrac{2\Phi}{c^2},~1-\dfrac{2\Phi}{c^2},~1-\dfrac{2\Phi}{c^2}\right)
\end{gather}

The gravitational potential due to the Milky Way galaxy is made up from the gravitational effects of the bulge, disk, and dark matter halo. The Miyamoto-Nagai disk, Hernquist bulge, and Navarro-Frenk-White potential models are used to model the gravitational influence of the galactic disk, bulge, and halo, respectively.

The potential of the Miyamoto-Nagai disk \citep{Miyamoto1975} is given by
\begin{gather}
\Phi_d=-\frac{GM_d}{\sqrt{x^2+y^2+\left(r_d+\sqrt{z^2+b_d^2}\right)^2}}
\label{diskpot}
\end{gather}
where $M_d=10\times10^{10}\,M_\odot$ is the mass of the disk, $r_d=6.5$\,kpc is the scale length of the disk, and $b_d=0.26$\,kpc is the scale height of the disk \citep{Nusser2009}.

The potential of the Hernquist Bulge \citep{Hernquist1990} is given by
\begin{gather}
\Phi_b=-\frac{GM_b}{\sqrt{x^2+y^2+z^2}+r_b}
\label{bulgepot}
\end{gather}
where $M_b=3.4\times10^{10}\,M_\odot$ is the mass of the bulge and $r_b=0.7$\,kpc is the scale length of the bulge.

The potential of the Navarro-Frenk-White Halo \citep{Navarro1997} is given by
\begin{gather}
\Phi_h=-\frac{GM_h}{\sqrt{x^2+y^2+z^2}}\ln\left(\frac{\sqrt{x^2+y^2+z^2}}{r_h}+1\right)
\label{halopot}
\end{gather}
where $M_h$ is the mass of the halo and $r_h$ is the scale length of the halo. The mass and scale lengths are calculated from the virial mass $M_v$ of the halo, which is the enclosed halo mass at the virial radius $R_v$. The exact size of a galaxy is difficult to quantify as the halo mass density extends out continuously into intergalactic space, and the virial radius can be thought of as the radius beyond which the halo blends into the background matter in the universe. For the Milky Way, the virial mass is roughly $M_v=150\times10^{10}\,M_\odot$ \citep{Dehnen2006}. The virial radius is calculated from the virial mass using \citep{Kafle2014}
\begin{gather}
R_v=\left(\frac{2M_vG}{H_0^2\Omega_m\Delta_{th}}\right)^{1/3}\approx294.5\,\mathrm{kpc}
\end{gather}
where $H_0=70.4\times10^{-3}\,\mathrm{km}\,\mathrm{s}^{-1}\,\mathrm{Mpc}^{-1}$ is the Hubble constant, $\Omega_m=0.3$, and $\Delta_{th}=340$ is the Hubble constant, matter density of the universe, and over-density of dark matter compared to the average matter density, respectively \citep{Bennett2013}.

The mass and scale lengths of the dark matter halo are related to the virial mass and virial radius via the dark matter halo concentration, which is described by the halo concentration parameter $c_h$, approximated by \citep{Bullock2005}
\begin{gather}
c_h\simeq9.6\left(\frac{M_v}{10^{13}M_\odot}\right)^{-0.13}(1+z)^{-1}\approx12
\end{gather}
where $z$ is the redshift, which is zero for host dark matter halos. The mass \citep{Naray2009} and scale length \citep{Kafle2012} of the halo are then given by 
\begin{gather}
M_h=\frac{M_v}{\ln(c_h+1)-\frac{c_h}{c_h+1}}\approx91.4\times10^{10}\,M_\odot\\
r_h=\frac{R_v}{c_h}\approx24.5\,\mathrm{kpc}
\end{gather}

The gravitational potential of the Milky Way is then the sum of each of the individual components.

\subsection{Relativistic Rocket}

For a relativistic rocket, the proper acceleration $a$ (i.e.~the acceleration as experienced by the traveller) is related to the rate of change of mass of the rocket \cite{Bade1953} by 
\begin{gather}
\frac{1}{c}\int_0^\tau a\,d\tau=-\frac{v_e}{c}\int_0^\tau \frac{1}{m}\frac{dm}{d\tau}\,d\tau
\end{gather}
where $v_e$ is the effective exhaust velocity of the propellants. If the rocket expends all its fuel after a proper time of $\tau_f$, then it is straightforward to show that
\begin{gather}
m_0 = m_r\exp\left(\dfrac{1}{v_e}\int_0^{\tau_f} a(\tau)\,d\tau\right)
\label{rockmass}
\end{gather}
where $m_r$ is the final mass of the rocket. If the mass of the fuel is $m_f$, then $m_0=m_f+m_r$, and hence
\begin{gather}
m_f=m_r\left[\exp\left(\dfrac{1}{v_e}\int_0^{\tau_f} a(\tau)\,d\tau\right)-1\right]=m_r\beta
\label{mfuelreq}
\end{gather}
where $\beta$ is the fuel-to-empty rocket mass ratio. Smaller values of $v_e$ will result in a larger value of $\beta$, and hence more fuel will be required as expected.

To compute the rocket trajectories around the Milky Way, we need to prescribe an acceleration four-vector for Equation \eqref{geoaccel}. Given the magnitude of the proper acceleration, $a$, we determine the components of the four-acceleration by setting the thrust vector. We let the spatial components of the four-acceleration be
\begin{gather}
a^i=(a^x,a^y,a^z)=a_s(a_x,a_y,a_z)
\end{gather}
where $a_x$, $a_y$, and $a_z$ are the components of a unit vector, so that
\begin{gather}
a^i=a_s(a_x\mathbf{\hat{x}}+a_y\mathbf{\hat{y}}+a_z\mathbf{\hat{z}})=|a^i|(a_x\mathbf{\hat{x}}+a_y\mathbf{\hat{y}}+a_z\mathbf{\hat{z}})
\end{gather}
and
\begin{gather}
a_x^2+a_y^2+a_z^2=1
\end{gather}
The values of $a^t$ and $a_s$ are determined via the normalisation conditions from Equations \eqref{norm1} and \eqref{norm2}, leading to 
\begin{gather}
a^t=-ak\frac{g_{ii}}{g_{tt}}\sqrt{\frac{g_{tt}}{g_{ii}[g_{ii}k^2+g_{tt}(u^t)^2]}}\\
a_s = au^t\sqrt{\frac{g_{tt}}{g_{ii}[g_{ii}k^2+g_{tt}(u^t)^2]}}
\end{gather}
where
\begin{gather}
k \equiv u^xa_x+u^ya_y+u^za_z
\end{gather}
and the metric terms are
\begin{gather}
g_{tt}=-c^2\left(1+\frac{2\Phi}{c^2}\right)\\
g_{ii}=1-\frac{2\Phi}{c^2}
\end{gather}

\subsection{Genetic Algorithms}
When approaching an optimisation problem, many types of optimisation methods can be employed, each offering unique advantages as well as disadvantages. \cite{Betts1998} explored various numerical optimisation methods commonly used in trajectory problems, and compared the benefits of each; \cite{Vinko2008} explored different optimisation methods to solve several spacecraft trajectory problems, and concluded that the most accurate solutions are produced by a combination of different solvers. 

Since the exact mass and size parameters of the Milky Way are still open to debate, a simple optimisation algorithm will suffice since we are only concerned with calculating an approximate optimal trajectory. A genetic algorithm was chosen due to its simplistic and heuristic nature, and are computationally inexpensive to run since they also do not depend on any derivatives and their respective matrices like most other optimisation methods. \cite{Charbonneau1995} explored the use of genetic algorithms in astronomy and astrophysics to solve a variety of problems, and demonstrates their simplicity and robustness when compared with conventional optimisation techniques. 

To simulate the genetic process, we start with an initial sample size and gradually evolve it toward the optimal solution \citep{Coley1999}:
\begin{enumerate*}
\item A random population is first constructed, representing the first generation of the species.
\item The fitness of each member of the species is evaluated, with the fittest member of the species is passed onto the next generation (elitism).
\item The rest of the population is created by selecting the fittest members of the first generation (selection) using a linear probability distribution, with breeding also occurring between the fittest members (crossover).
\item A mutation is randomly introduced into certain members to create genetic diversity.
\item Steps 2 to 5 are repeated for each subsequent generation.
\end{enumerate*}

\section{Method}
\subsection{Boundary Conditions}

The Sun orbits the galactic core of the Milky Way at a distance $r_0$ of roughly 8.5\,kpc, velocity $v_0$ of about 220\,km\,s$^{-1}$ \citep{Kerr1986}, and an orbital period of roughly 220\,Myr \citep{Seeds2015}. The Sun's orbit is roughly elliptical, and oscillates up and down relative to the galactic plane \citep{Moore2011}. We will assume that the Sun lies and stays within the galactic plane. 

The initial coordinates of the Sun are set at
\begin{gather}
x^\alpha=(0,0,-r_0,0)
\end{gather}
and the initial four-velocity is
\begin{gather}
u^\alpha=(u^t,-v_0,0,0)
\end{gather}
where $u^t$ is determined from Equation \eqref{norm1} to be 
\begin{gather}
u^t=\pm\sqrt{-\frac{c^2+g_{ii}[(u^x)^2+(u^y)^2+(u^z)^2]}{g_{tt}}}
\label{utinit}
\end{gather}

A rocket arriving at a galactic destination $(x_f,y_f,z_f)$ must also have the correct orbital velocity of the star system. Galactic rotation curves show that the orbital velocity as a function of radial distance is roughly constant outside the galactic bulge \citep{Jones2004}. For the Milky Way, the orbital velocity remains roughly constant around 220\,km\,s$^{-1}$ outside a radius of about 3\,kpc. Thus, the rocket must arrive at its final destination with a final speed $v_f$ of 220\,km\,s$^{-1}$, with four-velocity components
\begin{gather}
u_f^x=v_f\sin\phi_f\sin\theta_f\\
u_f^y=-v_f\cos\phi_f\sin\theta_f\\
u_f^z=-v_f\cos\theta_f
\end{gather}
where the final inclination angle $\theta_f$ and azimuthal angle $\phi_f$ are given by
\begin{gather}
\phi_f=\tan^{-1}\frac{y_f}{x_f}
\label{phifeqn}\\
\theta_f=\cos^{-1}\frac{z_f}{\sqrt{x_f^2+y_f^2+z_f^2}}
\label{thetafeqn}
\end{gather}

The equations of motion are integrated using Matlab's \texttt{ode45} solver, which is a non-stiff solver that utilises an explicit fifth-order Runge-Kutta method. 

The Christoffel symbols were calculated from an m-file developed by \cite{Engineering2014}, which is available on the MathWorks File Exchange website. The code was analysed before being extensively tested on some common spacetime metrics, and successfully reproduced their corresponding Christoffel symbols.

\subsection{Implementing the Genetic Algorithm}
The equations of motion require four input variables from the rocket: the magnitude $a$ of the four-acceleration, and the three components of the unit thrust vector. To reduce the number of variables that need to be solved, the unit thrust vector of the rocket is parameterised in spherical coordinates using the inclination $\theta$ and azimuthal $\phi$ angles, where $0\leq\theta\leq\pi$, $0\leq\phi\leq2\pi$, and
\begin{gather}
a_x = \sin\theta\cos\phi\\
a_y = \sin\theta\sin\phi\\
a_z=\cos\theta
\end{gather}
The three variables, $a$, $\theta$, and $\phi$ need to vary throughout the journey as a function of the proper time $\tau$. A spline is fitted through 2 splines points of each variable at $\tau=0$ and the final integration time $\tau_{\mathrm{end}}$. There is a total of 6 variables that need to be optimised by the genetic algorithm.

In the optimisation process, minimum and maximum bounds are specified for each variable $x_i$, denoted by $x_{i,\min}$ and $x_{i,\max}$, respectively. Each variable is represented as a binary string $X_i$ of length $\ell$, and hence each solution string is of length $n\ell$. Note that the maximum decimal value of the binary string is $2^\ell-1$. The initial sample of solutions is then obtained by randomly generating random integers between 0 and $2^\ell-1$, and the binary strings are then mapped to a particular value of each variable using:
\begin{gather}
x_i=x_{i,\min}+\frac{X_i}{2^\ell-1}(x_{i,\max}-x_{i,\min})
\end{gather}
 
\subsection{Parameters of the Genetic Algorithm}
The fitness function used to quantify each solution is
\begin{flalign}
f = & ~(x-x_f)^2+(y-y_f)^2+(z-z_f)^2\\
&+k_v[(u^x-u_f^x)^2+(u^y-u_f^y)^2+(u^z-u_f^z)^2]+k_\tau\tau^2
\end{flalign}
where $k_v$ and $k_\tau$ are the velocity and proper time weighting factors, respectively. The fitness for each solution sample is calculated at each point along the trajectory, and the minimum fitness along that path is taken to be the fitness for that solution.

The bounds for $a$, $\theta$, and $\phi$ are set to be
\begin{gather}
\begin{array}{ll}
a_{\min}=-10^{-5}\,\mathrm{g} & a_{\max}=+10^{-5}\,\mathrm{g}\\
\theta_{\min}=0 & \theta_{\max}=\pi\\
\phi_{\min}=0 & \phi_{\max}=2\pi\\
\end{array}
\end{gather}
where a negative acceleration is equivalent to an acceleration in the opposite direction. The acceleration magnitudes are sufficiently small enouugh to be classified as a low-acceleration rocket.  

The parameters of the genetic algorithm are shown in Table \ref{variousparam}. With 10 binary digits, the resolutions of each parameter are
\begin{gather}
a_{\mathrm{res}}=1.96\times10^{-9}\,\mathrm{g}\\
\theta_{\mathrm{res}}=0.18^\circ\\
\phi_{\mathrm{res}}=0.35^\circ
\end{gather}

\begin{table}[h!]
\begin{center}
\begin{tabular}{|c|c|}
\hline
Parameter & Value\\
\hline
Binary digits & 10\\
\hline
Sample size & 200\\
\hline
Generations & 100\\
\hline
Selection probability & 25\%\\
\hline
Mutation probability & 25\%\\
\hline
Spline points & 2\\
\hline
Velocity weighting factor & $10^{-5}$\\
\hline
Proper time weighting factor & $10^{-2}$\\
\hline
\end{tabular}
\caption{Parameters of the genetic algorithm used in the interstellar trajectory calculations.}
\label{variousparam}
\end{center}
\end{table}

\subsection{Rocket Parameters}
The rocket mass and exhaust velocity has a significant role when it comes to choosing the type of propulsion system used for an interstellar journey. The values used will be based on expected technologies available in the future. When travelling to distant star systems, interstellar journeys are likely to require a multi-generational spacecraft. The empty mass of the interstellar spacecraft used for the trajectory calculation is set to 1000 tons to account for a multi-generational journey. It should be noted that the actual empty mass of the spacecraft is not really important; rather, we are more concerned with the fuel-to-empty rocket mass ratio $\beta$. When calculating this ratio, three sources of propulsion will be considered: 
\begin{itemize*}
\item Fusion rocket: Fusion rockets utilise the process of nuclear fusion to propel the spacecraft forward. The high-velocity charged particle exhaust can be reflected from the fusion reaction using electromagnetic fields, resulting in effective exhaust velocities on the order of 10,000\,km\,s$^{-1}$ \citep{Matloff2010}. The fusion rocket here has an effective exhaust velocity of 10,000\,km\,s$^{-1}$.
\item Antimatter rocket: Proton-antiproton annihilation is extremely efficient, and can convert more than 50\% of the fuel mass to usable exhaust kinetic energy. \citep{Westmoreland2010} showed that an effective exhaust velocity of up to $0.58c$ can be achieved by using an electromagnetic field to collimate the charged pion products into an exhaust jet. For the antimatter drive used here, an effective exhaust velocity of $0.5c$ is used for the reaction products of the proton-antiproton reaction.
\item Photon rocket: A photon rocket is an ideal rocket that generates thrust by emitting photons generated through electron-positron annihilation\footnote{While photon rockets are a type of antimatter rocket, we will use the term ``antimatter rocket'' to refer to proton-antiproton annihilation, and ``photon rocket'' to refer to electron-positron annihilation.}, and thus has an exhaust velocity equal to the speed of light \citep{Tinder2006}.
\end{itemize*}

\subsection{Galactic Destinations}

Various final destinations around the Milky Way are considered:\\[0.3cm]
\indent A. Star systems located within the galactic plane.\\[0.1cm]
\indent B. Hypervelocity stars in the galactic plane.\\[0.1cm]
\indent C. Star systems in the galactic halo.\\

For each of the above scenarios, two different locations are chosen:
\begin{enumerate*}
\item Located in the 2nd quadrant.
\item Located in the 4th quadrant, requiring the rocket to perform a complete turnaround before aligning itself with the velocity vector of the star.  
\end{enumerate*} 

For each destination, we aim to arrive within 0.2\,kpc and within 20\,km\,s$^{-1}$ of the target destination. Note that star systems further away from the galactic center will take longer to complete their orbit, and hence their distance from the Sun could change appreciably with time. We have assumed that upon arrival, the location of each star system remains unchanged relative to the Sun. This assumption is reasonable as long as the time taken to reach the destination is significantly shorter than the period of the orbit. An integration period of 1\,Myr is used in the calculations, which is a relatively short time span in galactic orbits (recall that the period of the Sun is about 220\,Myr).

We will assume that star systems in the galactic disk lie within the galactic plane, which is a reasonable assumption since the average width of the disk is about 1\% of the diameter of the disk. Thus, we can further simplify the analysis by setting $\theta=\pi/2$ throughout the calculation, so that there are only 4 variables that need to be optimised for. To ensure that the inclination angle remains unchanged, the lower and upper bounds are set to $\pi/2$.

\section{Results}
For each scenario, the genetic algorithm is run 20 times, and the solution with the lowest final fitness is presented. The final position deviation, final three velocity deviation, proper time taken, maximum velocity reached, and fuel-to-empty rocket mass ratios for each propulsion system, are shown in Table \ref{summary}.

\subsection{Destination A1}

\begin{figure}[h!]
\centering
\includegraphics[trim = 130mm 0mm 130mm 0mm, clip, width=12cm]{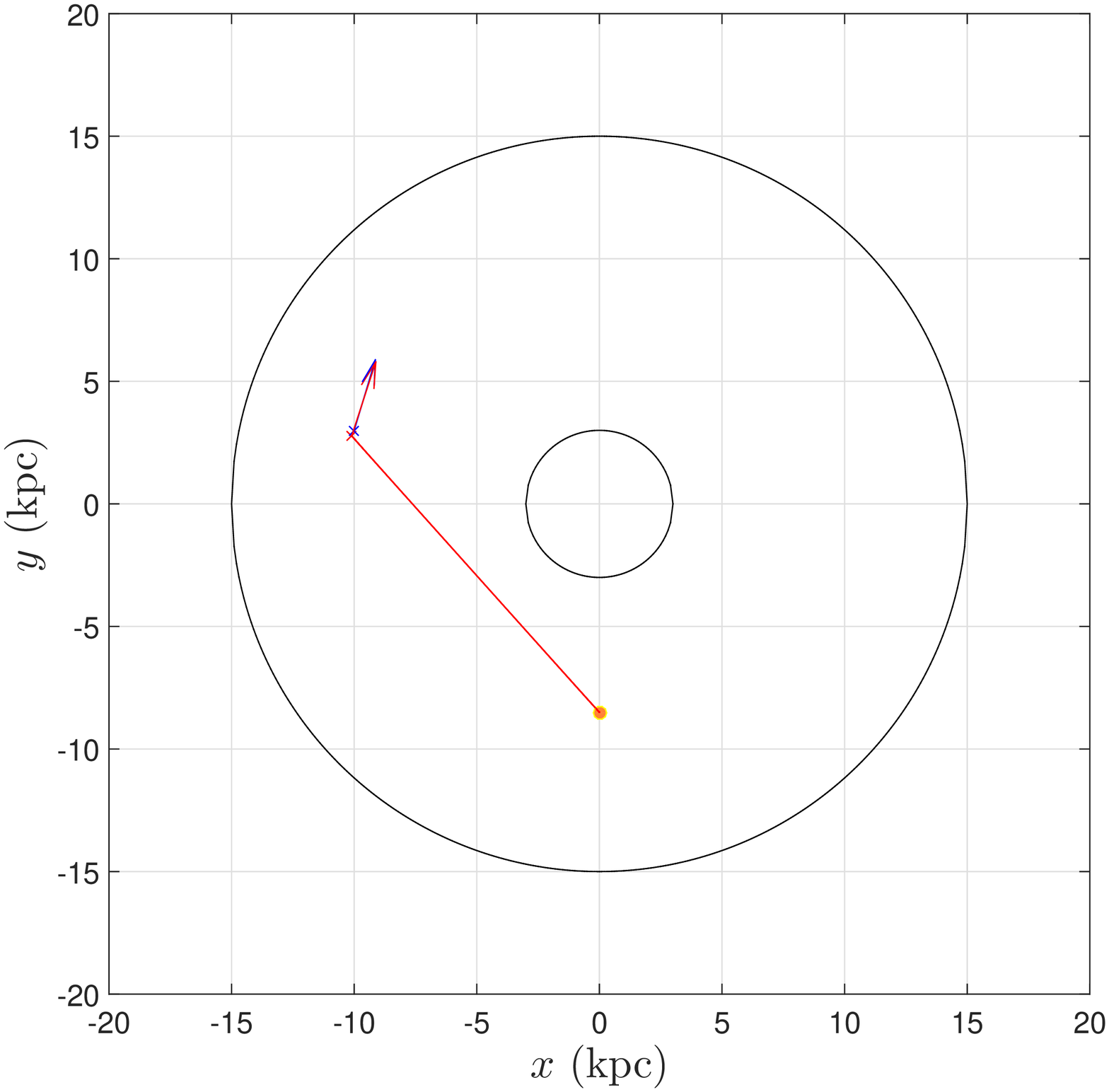}
\caption{Optimal trajectory (solid red) to destination A1. The location of the Sun is indicated by the orange dot, and the target and calculated final destinations are indicated by the red and blue cross, respectively. The red and blue arrows represent the calculated and required final velocity vector, respectively.}
\label{A1_1}
\end{figure}

The first star system is located within the galactic plane at $(-10,3,0)$, about 15\,kpc from the Sun. The results are shown in Fig.~\ref{A1_1} and Fig.~\ref{A1_2}.

\begin{figure}[h!]
\centering
\includegraphics[trim = 20mm 0mm 20mm 0mm, clip, width=12cm]{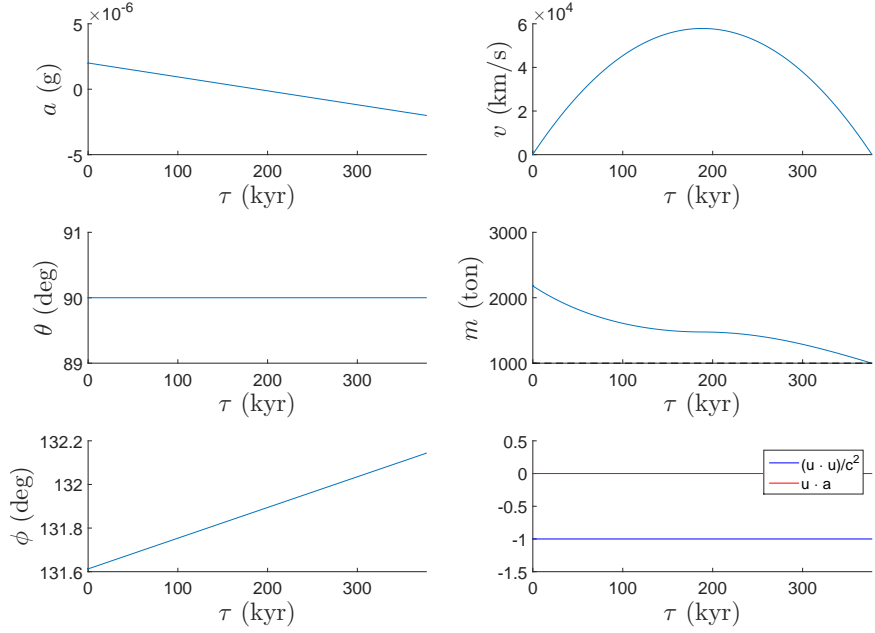}
\caption{Parameters of the optimal trajectory to destination A1.  The left column plots the variables $a$, $\theta$, and $\phi$ during the journey. The right column plots the velocity function (top), mass function (middle), and normalisation values (bottom). The mass functions shown in each scenario are for the antimatter rocket.}
\label{A1_2}
\end{figure}

\subsection{Destination A2}
\begin{figure}[h!]
\centering
\includegraphics[trim = 130mm 0mm 130mm 0mm, clip, width=12cm]{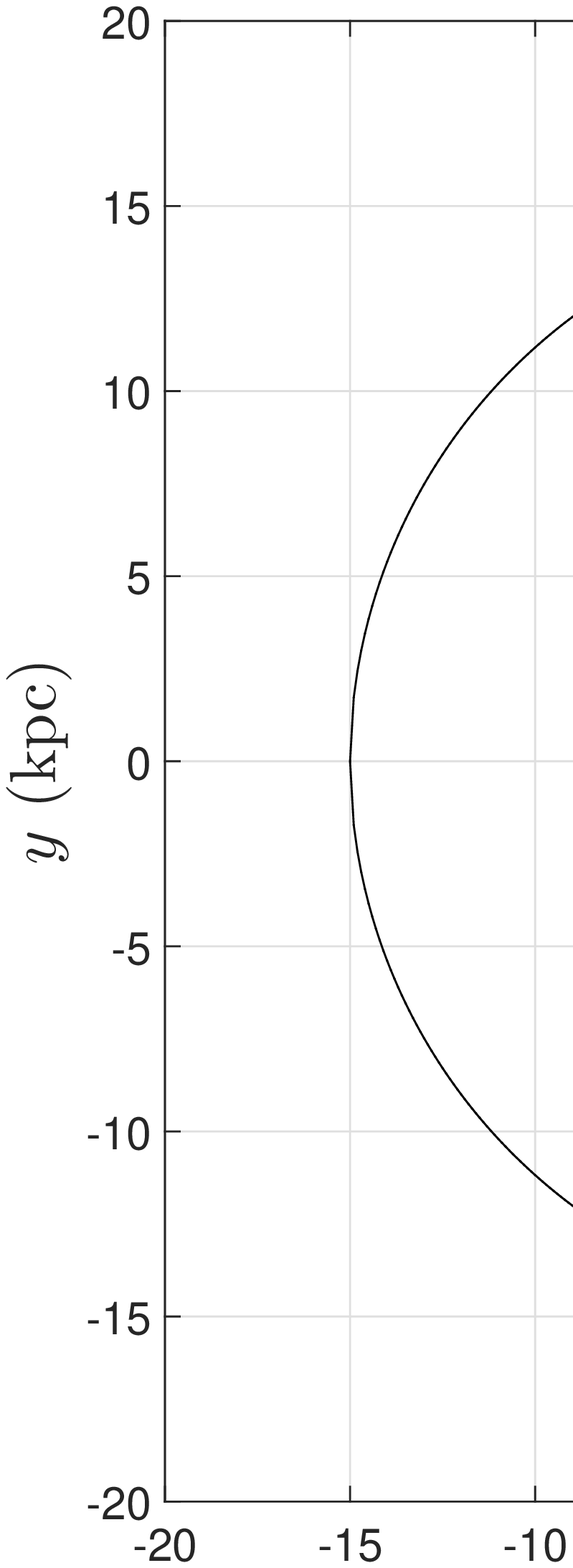}
\caption{Optimal trajectory to destination A2.}
\label{A2_1}
\end{figure}

The second star system is located within the galactic plane at $(8,-2,0)$, about 10\,kpc from the Sun. The results are shown in Fig.~\ref{A2_1} to \ref{A1_2}.

\begin{figure}[h!]
\centering
\subfigure{
\includegraphics[trim = 120mm 0mm 120mm 0mm, clip, width=5.5cm]{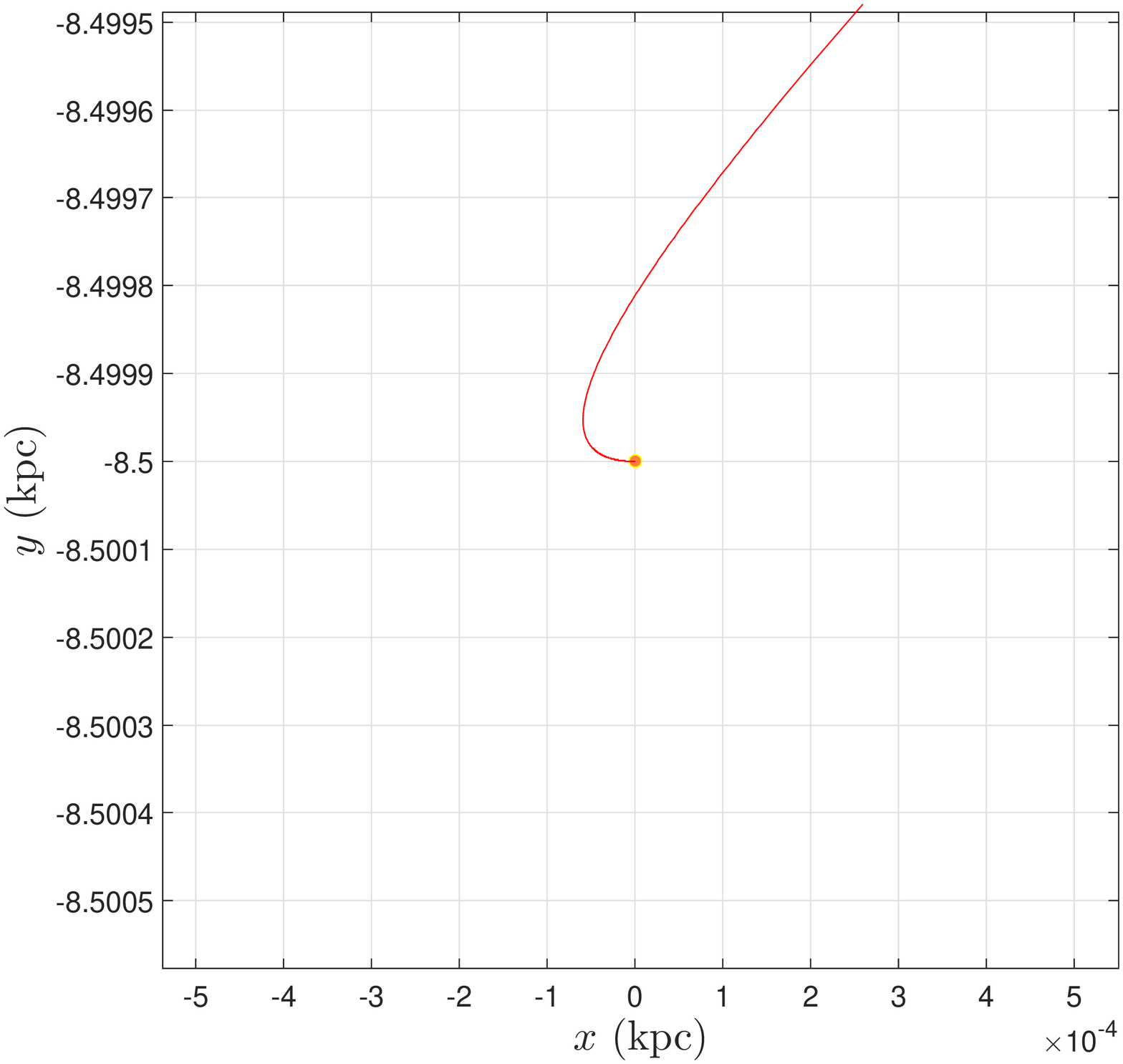}
}
\subfigure{
\includegraphics[trim = 120mm 0mm 120mm 0mm, clip, width=5.5cm]{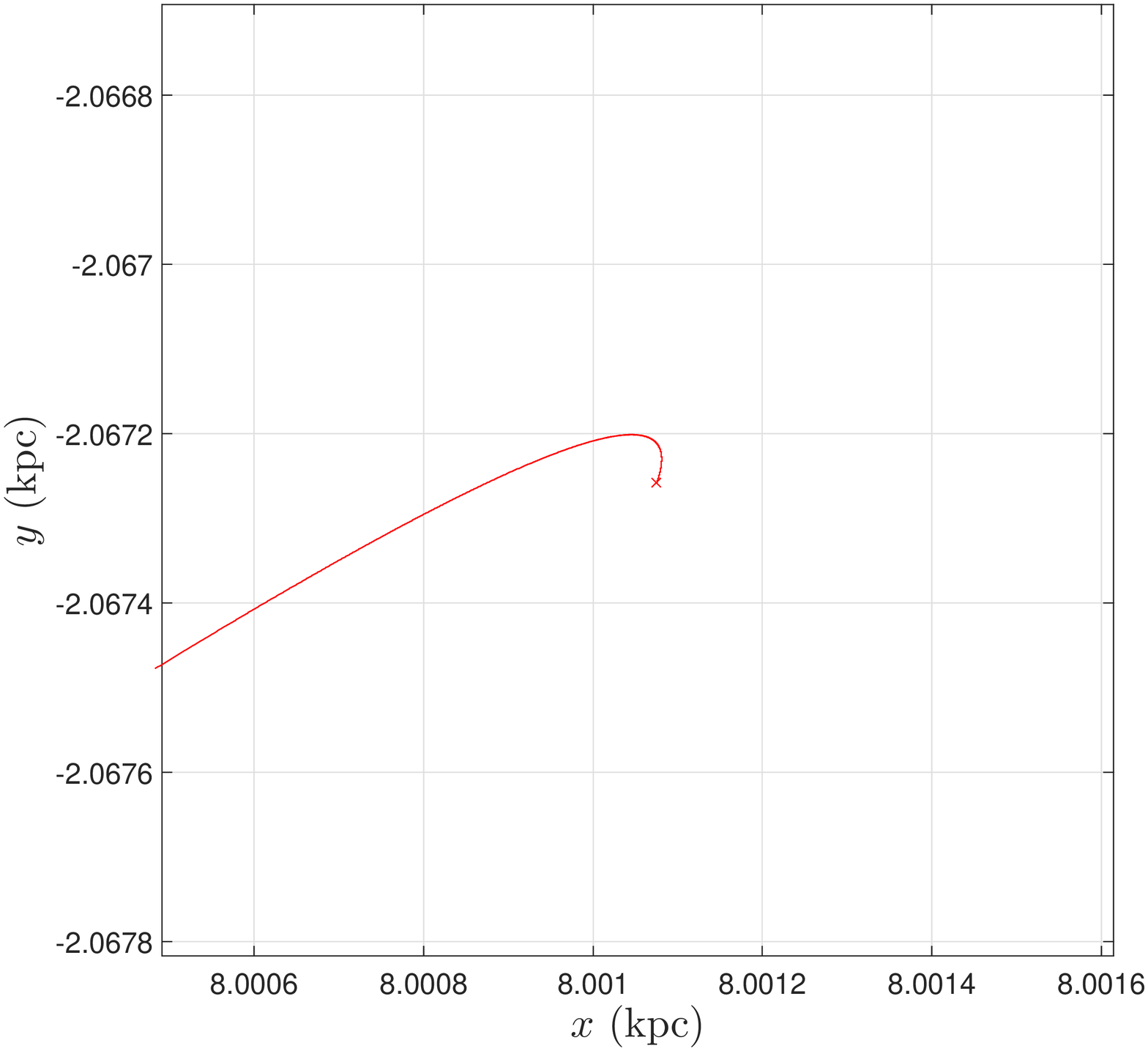}
}
\caption{Close up view of the optimal trajectory to destination A2 at the start (left) and end (right) points.}
\label{A3}
\end{figure}

\begin{figure}[h!]
\centering
\includegraphics[trim = 20mm 0mm 20mm 0mm, clip, width=12cm]{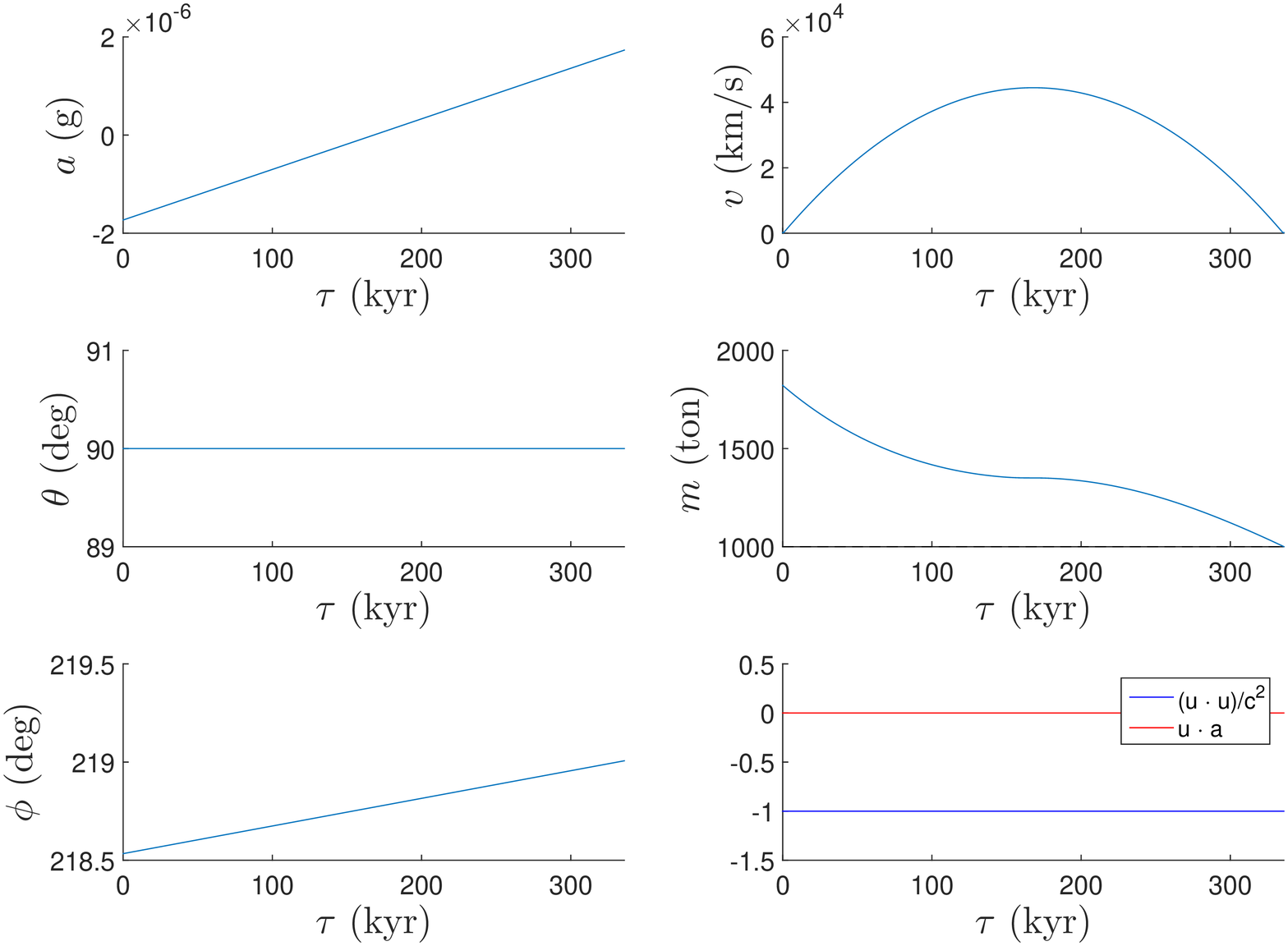}
\caption{Parameters of the optimal trajectory to destination A2.}
\label{A1_2}
\end{figure}

\subsection{Destination B1}
Hypervelocity stars (HVS) are stars with abnormally large velocities that exceed the escape velocity of the galaxy, travelling on radial paths with velocities exceeding 1000\,km\,s$^{-1}$ \citep{Baumgardt2006}. These stars are believed to originate from the core of galaxies, where supermassive blackholes are thought to exist \citep{Hills1988}. Reaching a hypervelocity star is of interest if one wishes to leave the galaxy, as orbiting around these stars provides a means of collecting energy during the journey. For a star on a radial trajectory, the velocity components are
\begin{gather}
u_f^x=v_f\cos\phi_f\sin\theta_f\\
u_f^y=v_f\sin\phi_f\sin\theta_f\\
u_f^z=v_f\cos\theta_f
\end{gather}
where the final angular coordinates are given by Equation \eqref{phifeqn} and \eqref{thetafeqn}.

\begin{figure}[h!]
\centering
\includegraphics[trim = 130mm 0mm 130mm 0mm, clip, width=12cm]{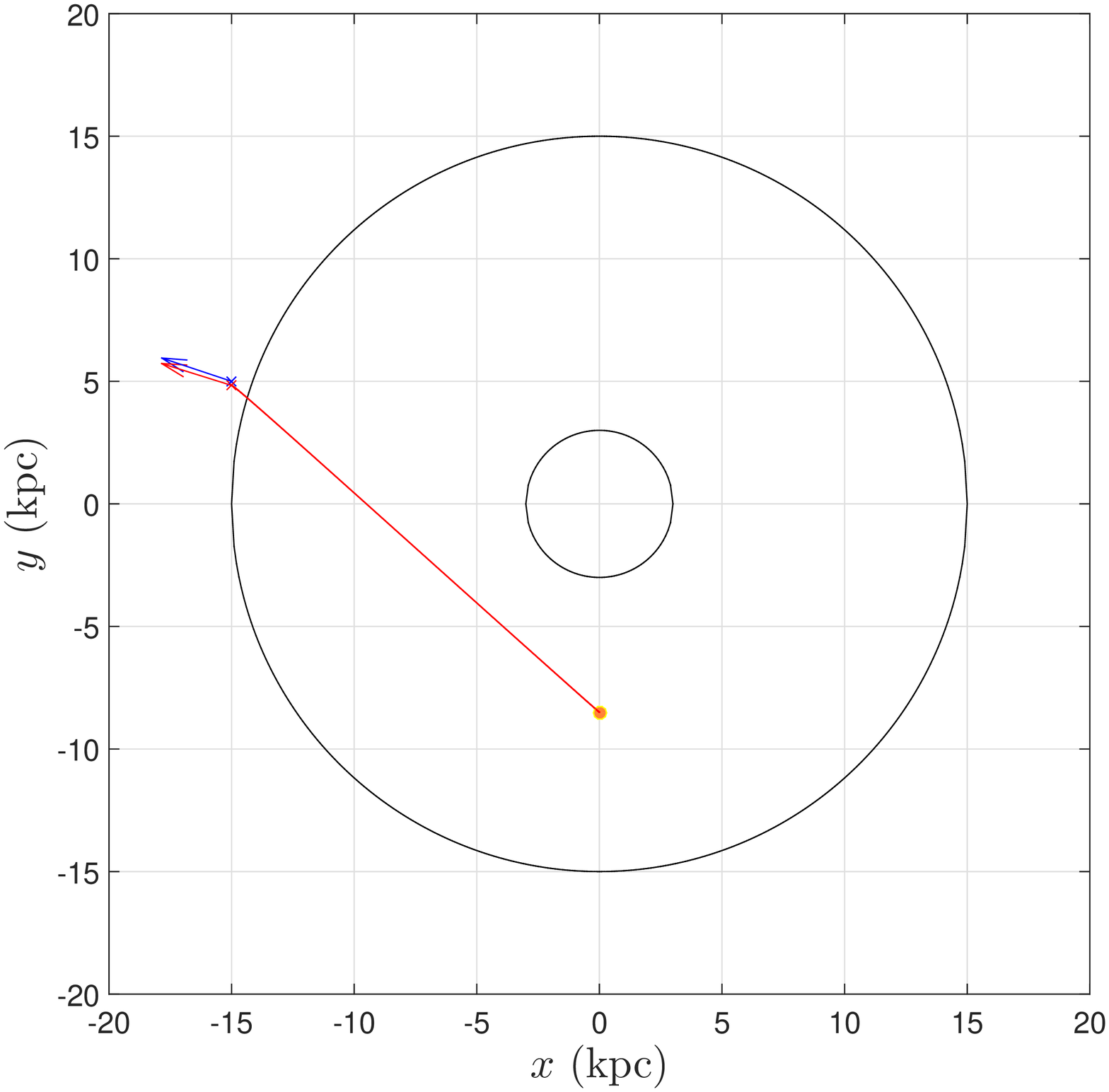}
\caption{Optimal trajectory to destination B1.}
\label{B1_1}
\end{figure}

\begin{figure}[h!]
\centering
\includegraphics[trim = 20mm 0mm 20mm 0mm, clip, width=12cm]{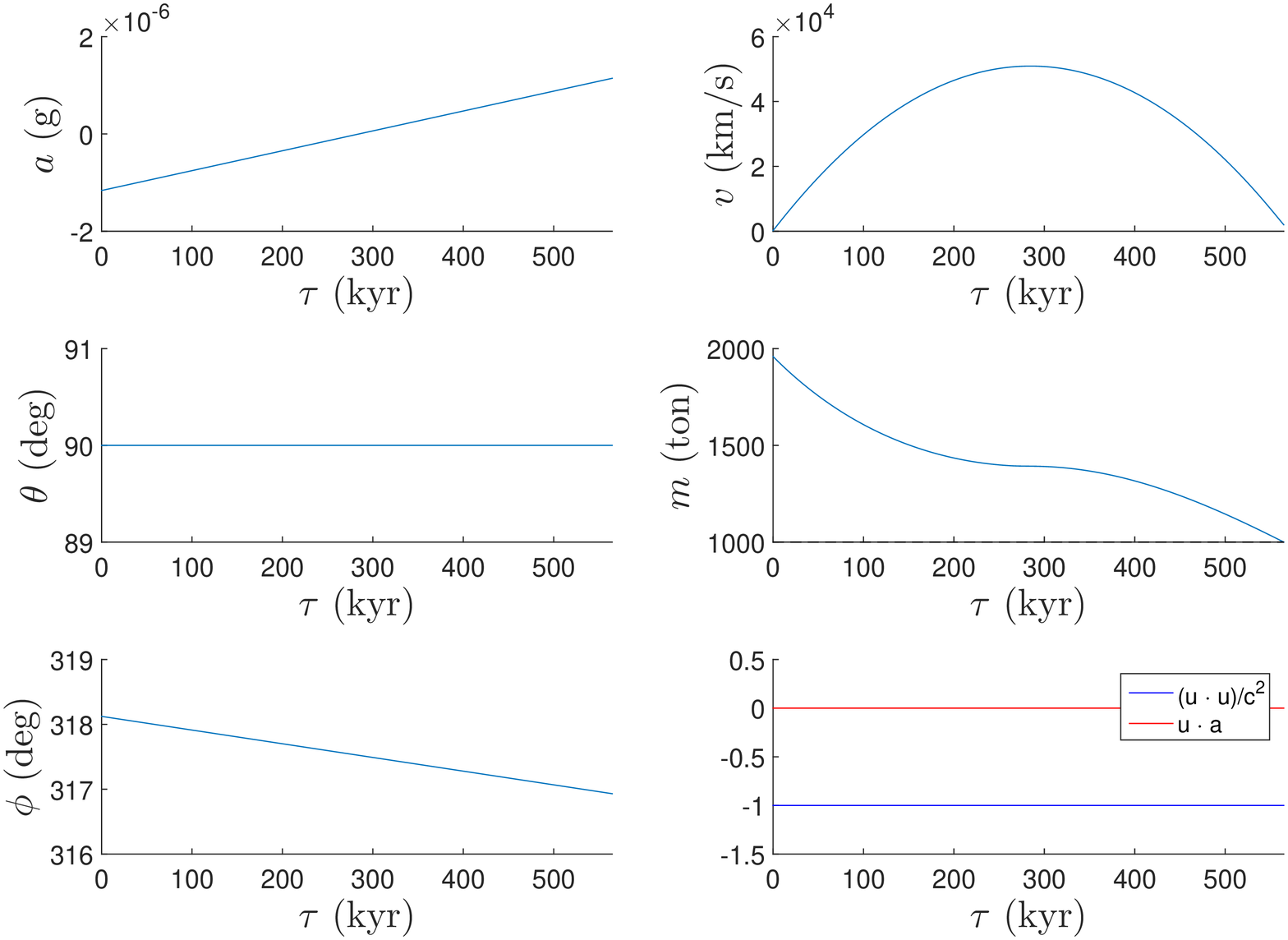}
\caption{Parameters of the optimal trajectory to destination B1.}
\label{B1_2}
\end{figure}

The first HVS is located at $(-15,5,0)$, about 20\,kpc from the Sun, and with a velocity magnitude of 2000\,km\,s$^{-1}$. The results are shown in Fig.~\ref{B1_1} and \ref{B1_2}.

\subsection{Destination B2}
The second HVS is located at $(12,-1,0)$, about 14\,kpc from the Sun, and with a velocity magnitude of 2000\,km\,s$^{-1}$. The results are shown in Fig.~\ref{B1_1} and \ref{B1_2}. 

\begin{figure}[h!]
\centering
\includegraphics[trim = 130mm 0mm 130mm 0mm, clip, width=12cm]{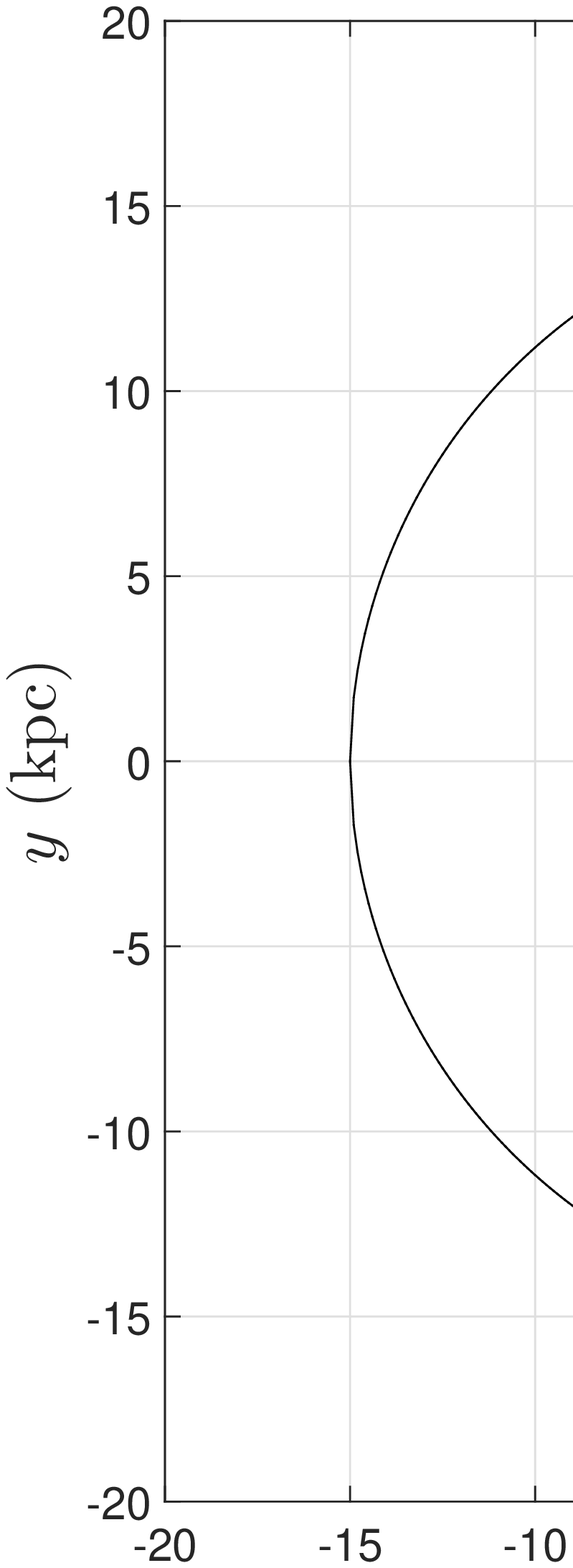}
\caption{Optimal trajectory to destination B2.}
\label{B2_1}
\end{figure}

\begin{figure}[h!]
\centering
\includegraphics[trim = 20mm 0mm 20mm 0mm, clip, width=12cm]{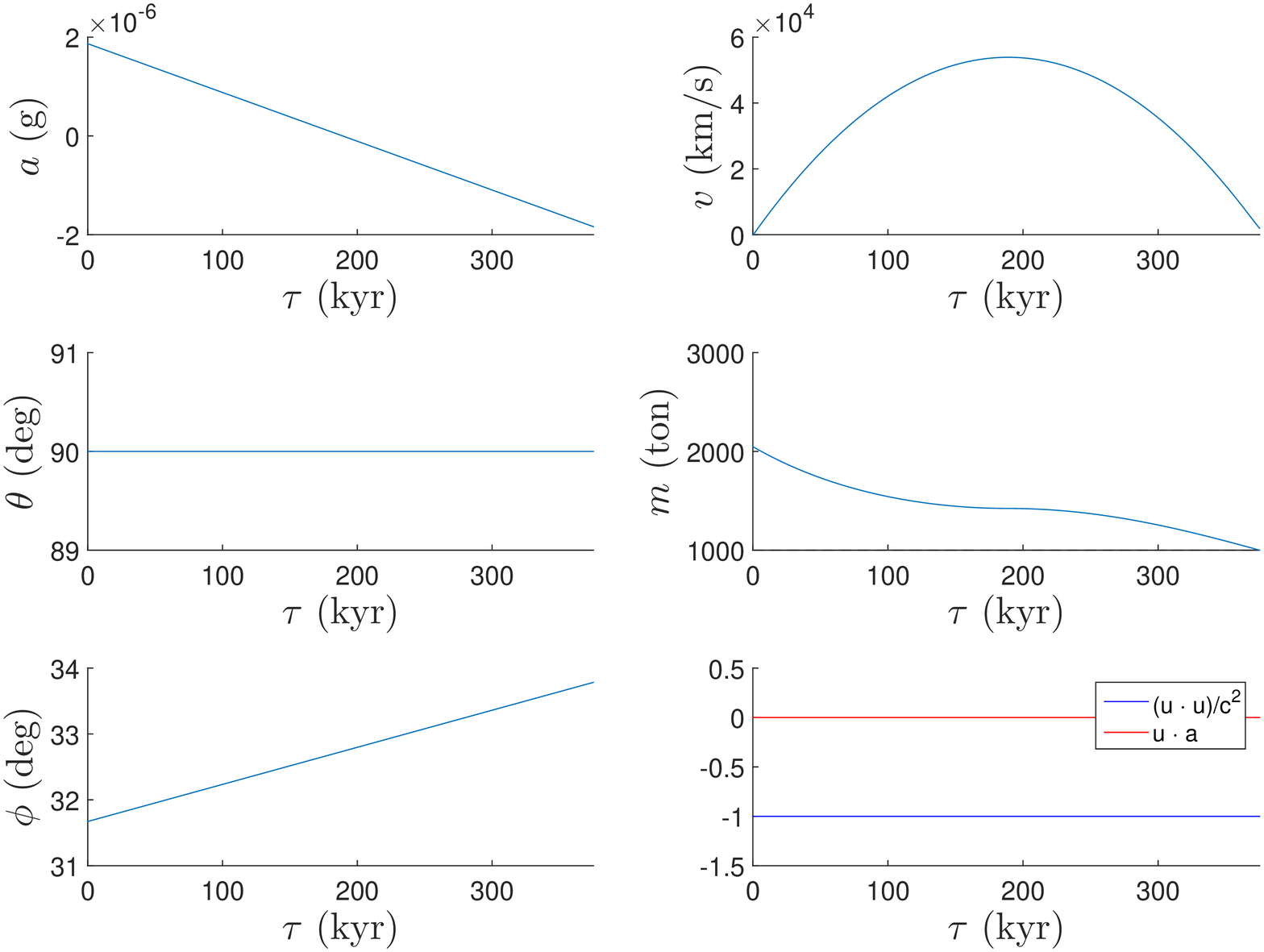}
\caption{Parameters of the optimal trajectory to destination B2.}
\label{B2_2}
\end{figure}

\subsection{Destination C1}
\begin{figure}[h!]
\centering
\includegraphics[trim = 20mm 0mm 20mm 0mm, clip, width=12cm]{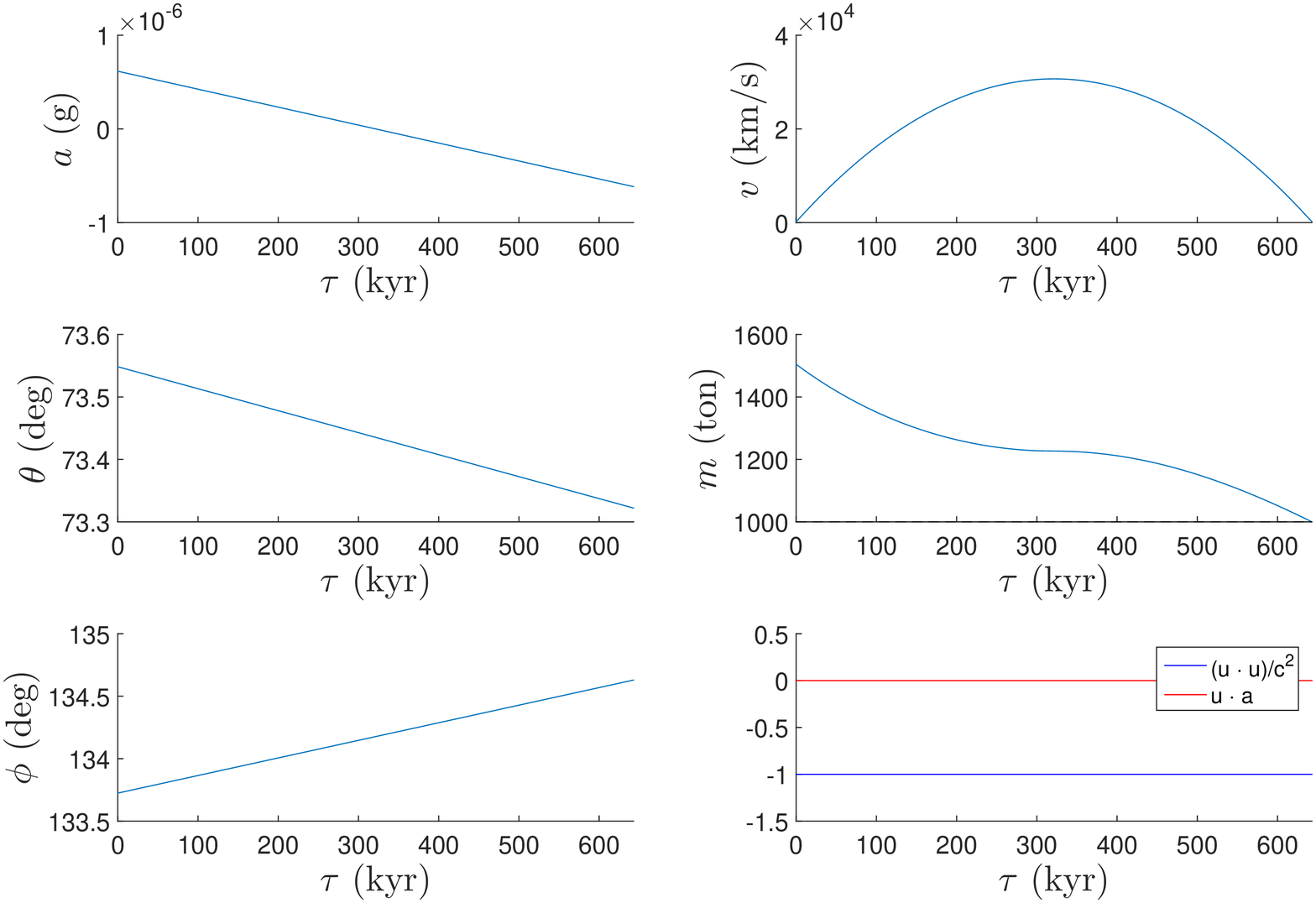}
\caption{Parameters of the optimal trajectory to destination C1.}
\label{C1_2}
\end{figure}

The next star system chosen will lie in the galactic halo to determine if the genetic algorithm can solve for the extra angular variable $\theta$. The star system chosen is located at $(-9,1,4)$, about 14\,kpc from the Sun. The results are shown in Fig.~\ref{C1_1} to \ref{C1_2}.

\begin{figure}[h!]
\centering
\includegraphics[trim = 130mm 0mm 130mm 0mm, clip, width=12cm]{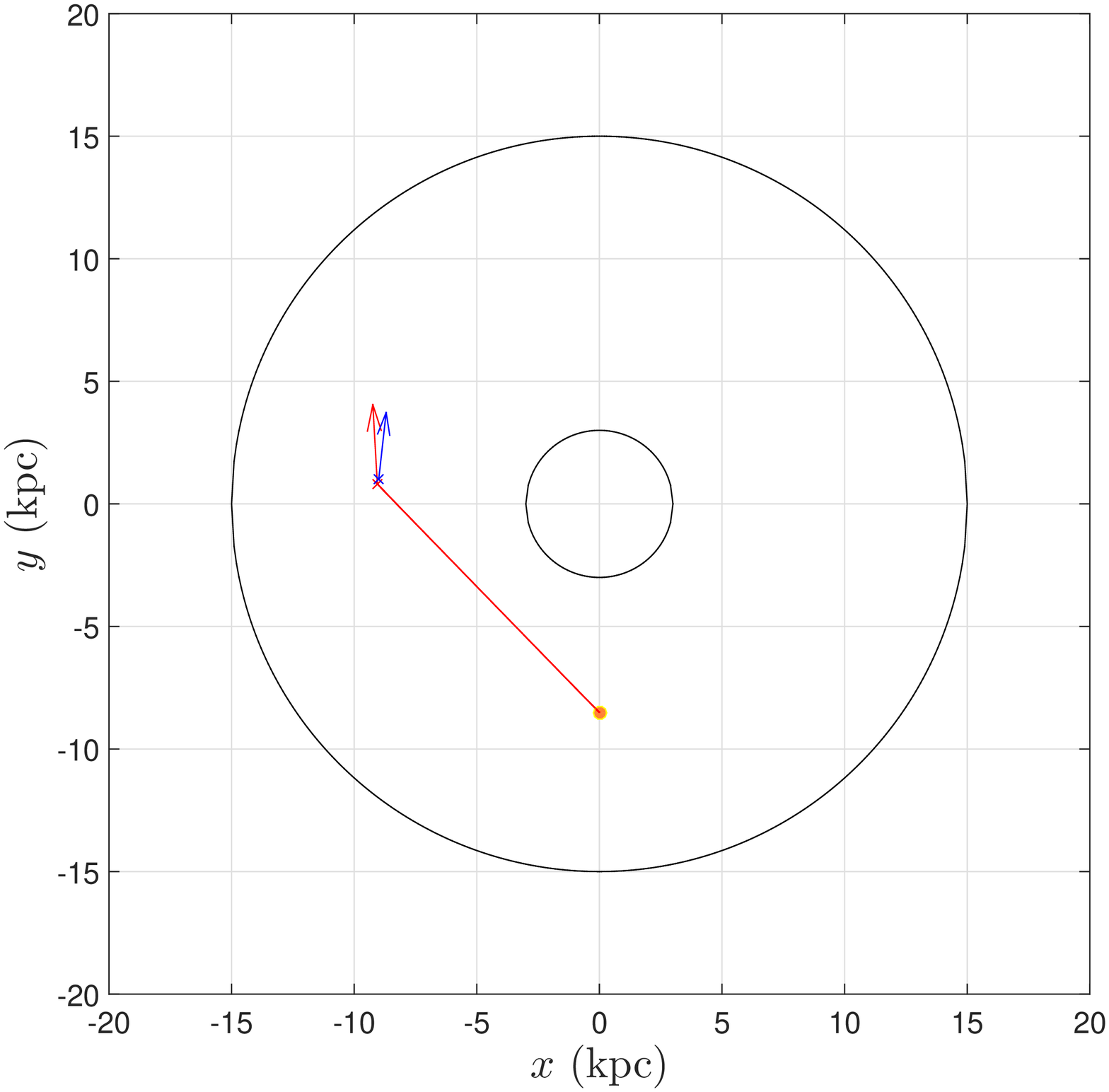}
\caption{Optimal trajectory to destination C1.}
\label{C1_1}
\end{figure}

\begin{figure}[h!]
\centering
\includegraphics[trim = 130mm 0mm 130mm 0mm, clip, width=12cm]{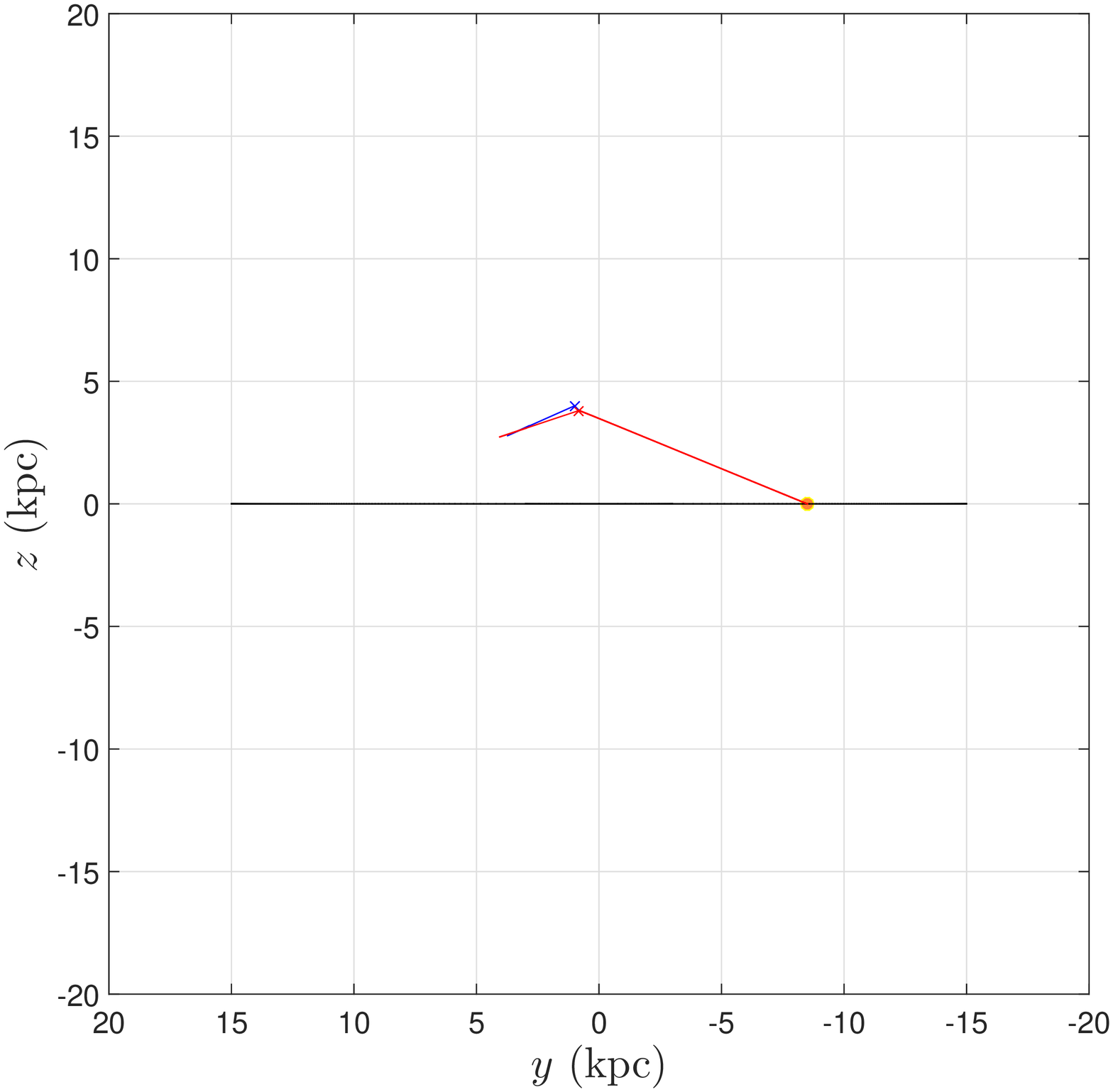}
\caption{Side view of optimal trajectory to destination C1.}
\label{C1_1}
\end{figure}

\subsection{Destination C2}
\begin{figure}[h!]
\centering
\includegraphics[trim = 10mm 0mm 10mm 0mm, clip, width=12cm]{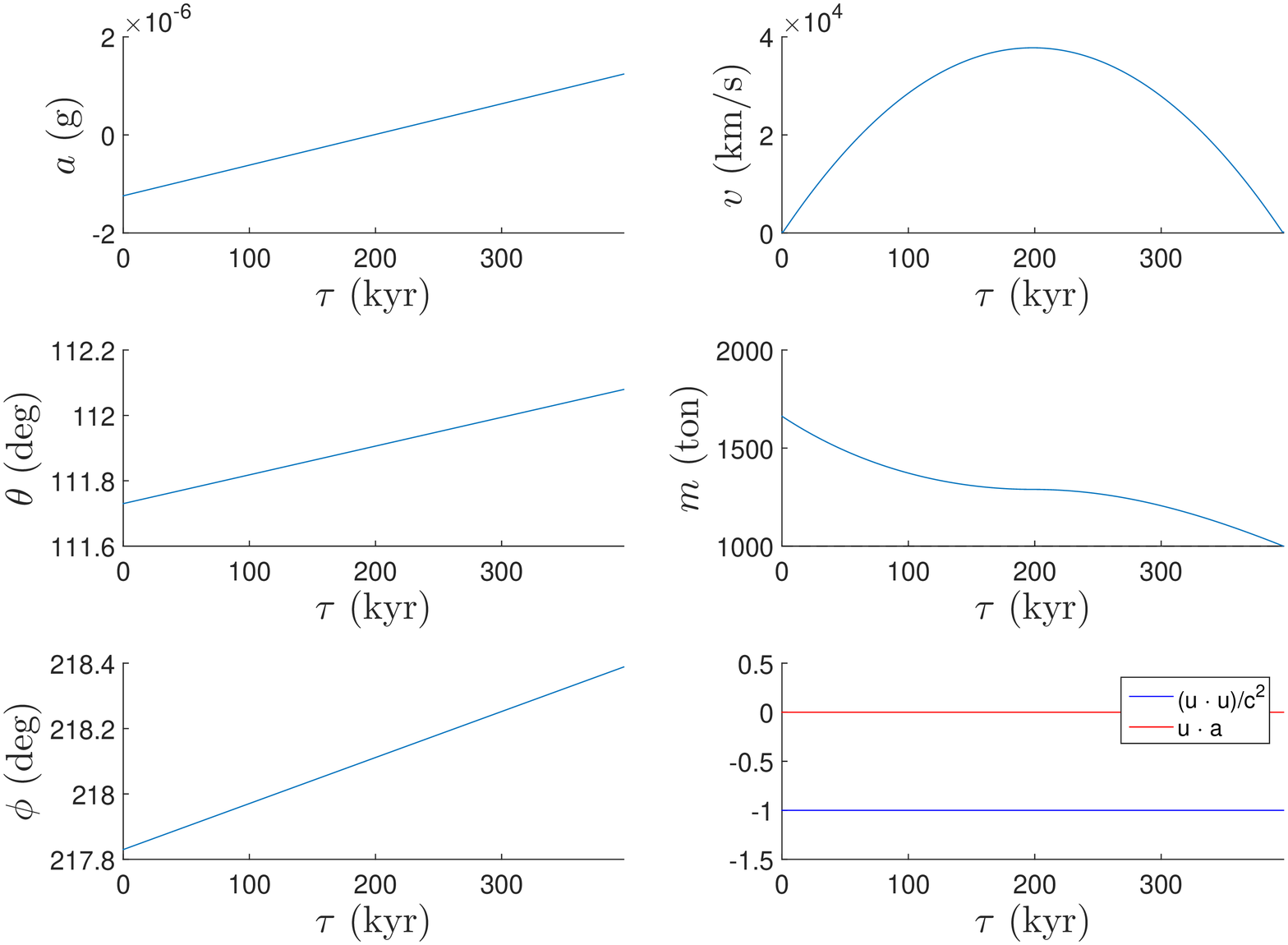}
\caption{Parameters of the optimal trajectory to destination C2.}
\label{C2_2}
\end{figure}

The final star system in the galactic halo is located at $(8,-2,3)$, about 11\,kpc from the Sun. The results are shown in Fig.~\ref{C2_1} to \ref{C2_2}. 

\begin{figure}[h!]
\centering
\includegraphics[trim = 130mm 0mm 130mm 0mm, clip, width=12cm]{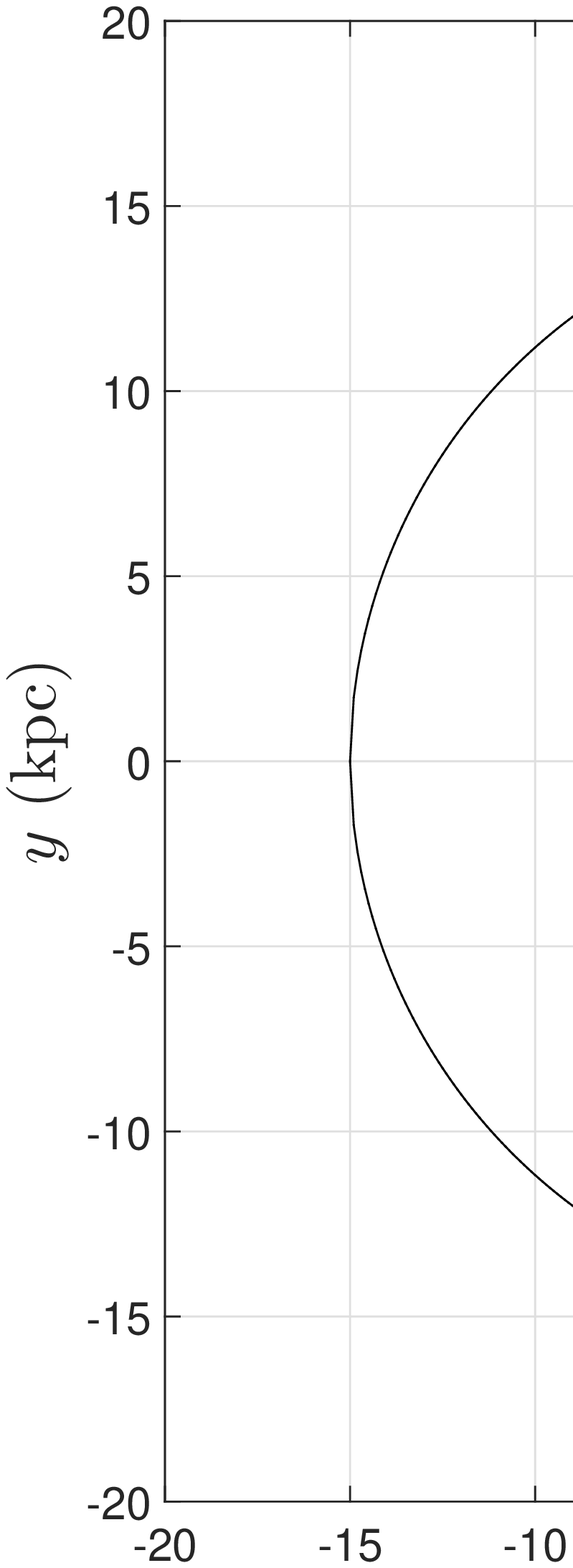}
\caption{Optimal trajectory to destination C2.}
\label{C2_1}
\end{figure}

\begin{figure}[h!]
\centering
\includegraphics[trim = 130mm 0mm 130mm 0mm, clip, width=12cm]{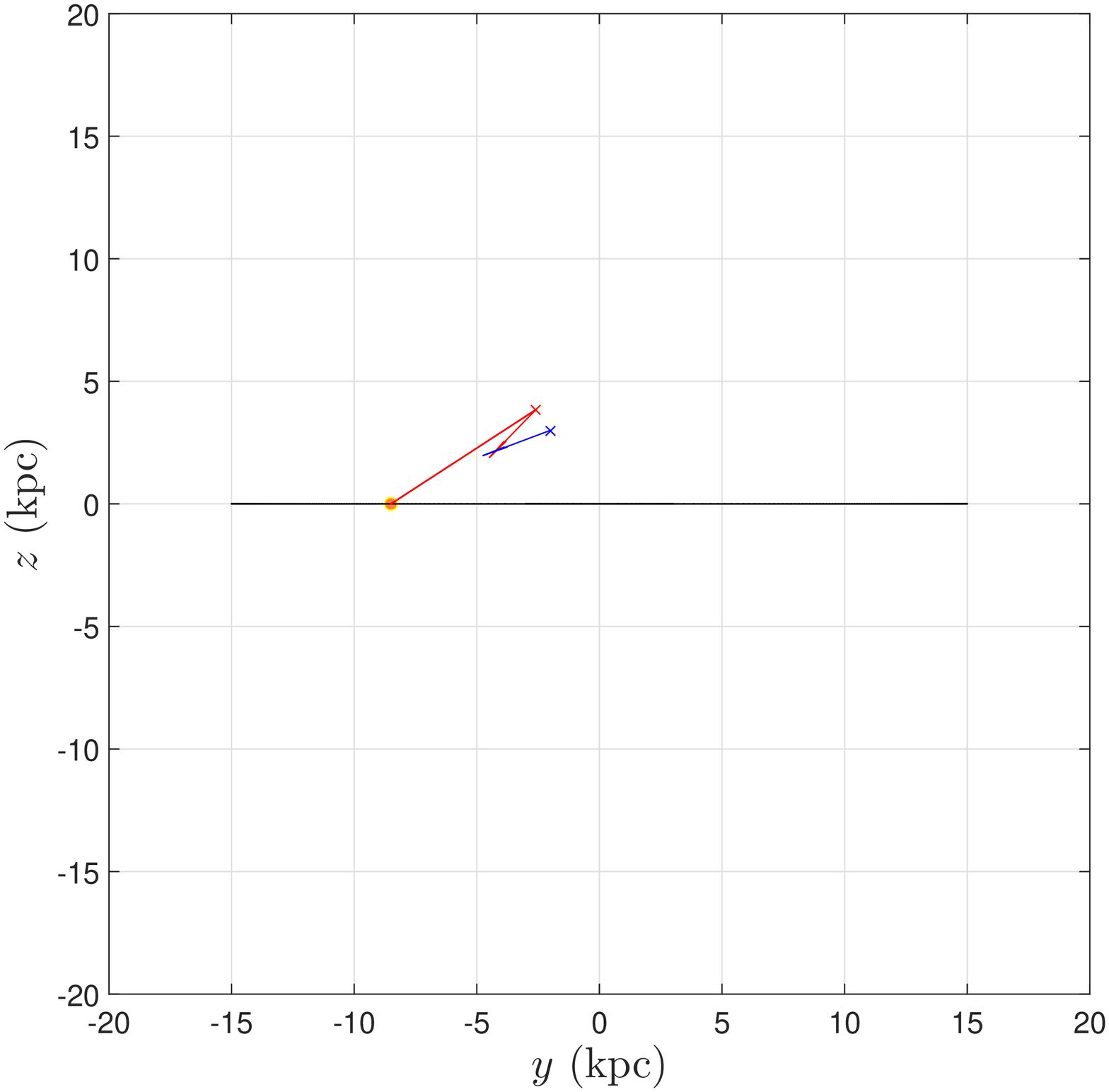}
\caption{Side view of optimal trajectory to destination C2.}
\label{C2_1}
\end{figure}

\section{Discussion}
A summary of the results is shown in Table \ref{summary}. For destinations within the galactic plane, the genetic algorithm is able to calculate optimal trajectories that arrive within 0.25\,kpc and 35\,km\,s$^{-1}$ of the desired star system. However, the algorithm struggles to find high quality solutions for destinations outside the galactic plane. A larger sample size and/or a higher number of generations are likely required to generate more accurate solutions to these destinations. The number of binary digits seems to be sufficient in each of the cases considered, and do not need to be increased unless one wishes to further refine a particular solution. For 200 samples and 100 generations, about 60\,mins was required for each application of the algorithm using a quad-core, 2.6\,GHz CPU with 8\,GB of RAM.

\begin{table}[h!]
\begin{center}
\begin{tabular}{|c|c|c|c|c|c|c|}
\hline
Destination & A1 & A2 & B1 & B2 & C1 & C2\\
\hline
Distance (kpc) & 15 & 10 & 20 & 14 & 14 & 11\\
\hline
Distance deviation (kpc) & 0.24 & 0.07 & 0.17 & 0.08 & 0.27 & 1.14\\
\hline
Velocity deviation (km\,s$^{-1}$) & 12 & 5 & 1 & 21 & 30 & 22\\
\hline
Proper time taken (kyr) & 377 & 336 & 565 & 375 & 643 & 397\\
\hline
Maximum velocity ($c$) & 0.19 & 0.15 & 0.17 & 0.18 & 0.10 & 0.13\\
\hline
$\beta_\mathrm{fusion}$ & $10^5$ & $10^4$ & $10^4$ & $10^4$ & $10^2$ & $10^3$\\
\hline
$\beta_\mathrm{antimatter}$ & 1.18 & 0.82 & 0.96 & 1.05 & 0.5 & 0.66\\
\hline
$\beta_\mathrm{photon}$ & 0.48 & 0.35 & 0.24 & 0.43 & 0.23 & 0.29\\
\hline
\end{tabular}
\caption{Summary of results.}
\label{summary}
\end{center}
\end{table}

Inspection of the normalisation values in each of the scenarios shows that the Matlab ODE solver calculated the variables to a high degree of accuracy. The linear acceleration profile also leads to a parabolic velocity profile, which in turn leads to the consistent mass function curves as seen above.

The use of the various propulsion systems clearly has a significant impact on the fuel-to-empty mass ratio $\beta$. For a low-acceleration rocket with a linear acceleration profile, the use of a photon drive results in $\beta$ values below 0.50, with values as low as 0.24. An antimatter drive produces $\beta$ values between 0.50 and 1.2, though most stay below 1.0. These values are similar to the chemical rockets in use today. The $\beta$ values of the fusion drive are on the order of $10^2$, and reach as high as $10^5$. Clearly, a fusion drive would not be practical for such journeys. An antimatter drive would likely be the minimum technological requirement for an interstellar journey. This would apply for even high acceleration journeys, since high accelerations generally require high rates of fuel consumption, which in turn results in a heavier spacecraft. However, high accelerations require shorter travel times, and this trade-off is one that is frequently encountered in trajectory optimisation problems.

Upon inspection of each of the trajectories, it is evident that the gravitational potential has minimal effect on the trajectory of a low-acceleration rocket.
In each case, the maximum velocity reached is between $0.1c$ and $0.2c$. While these speeds may appear large, they are still non-relativistic, with corresponding Lorentz factors of 1.005 and 1.021 respectively. At these speeds, roughly 350-400\,kyr was required to reach most of the destinations (with the exceptions being Locations B2 and C2). Clearly, this is not very practical for manned interstellar travel. To achieve more practical solutions, the maximum allowable acceleration magnitude would need to be increased. However, doing so raises many issues that would need to be addressed, and these are discussed in the final chapter.

\section{Conclusion}
The results of this research have demonstrated the use of genetic algorithms to optimise interstellar trajectories using a low-acceleration rocket. By varying the magnitude and orientation of the acceleration vector, we have calculated optimal paths to several galactic destinations. The algorithm showed remarkable success for destinations within the galactic plane, but produced less successful results for destinations outside the galactic plane. To produce more accurate trajectories to destinations outside the galactic plane, the sample size and/or number of generations would need to be increased.

For the mass calculations, three types of propulsive systems were considered: the fusion drive, antimatter drive, and the photon drive. The fuel-to-empty rocket mass ratio for each of these were calculated, and it was concluded that for a linear acceleration profile, the antimatter drive would represent the minimum technological requirement when undertaking an interstellar journey. The calculated values are conservative since low accelerations generally require low fuel consumption. However, low accelerations also require longer travel times; for each interstellar destination, low-acceleration rockets required several hundreds of thousands of years to reach their destination. Clearly, galactic colonisation is highly unrealistic using low-acceleration rockets.

Evidently, human expansion into the cosmos is extremely unlikely unless high acceleration rockets were used. However, higher accelerations often require unrealistic amounts of fuel, and this trade-off between travel time and fuel consumption is one that needs to be finely balanced. It remains to be seen whether anti-matter drives and photon drives are still viable candidates for these high acceleration journeys. Even if they were, the human race does not currently possess the technological capacity to manufacture antimatter on an industrial scale, and will unlikely attain the ability to do so for many more decades if not centuries to come. Consequently, the idea of using antimatter as a source of fuel, and hence the possibility of interstellar travel, will remain a scientific dream for the foreseeable future. If it is found that prohibitively large amounts of antimatter fuel are required, then mankind will probably need to turn to unconventional propulsion technologies, such as warp drives, for galactic colonisation.

\section{Acknowledgements}
The author would like to acknowledge the Australian Postgraduate Awards (APA), through which this work was financially supported.

\section*{References}

\bibliography{mybibfile}

\begin{thebibliography}{10}
\expandafter\ifx\csname url\endcsname\relax
  \def\url#1{\texttt{#1}}\fi
\expandafter\ifx\csname urlprefix\endcsname\relax\def\urlprefix{URL }\fi
\expandafter\ifx\csname href\endcsname\relax
  \def\href#1#2{#2} \def\path#1{#1}\fi

\bibitem{Cassenti1997}
B.~Cassenti, {Optimisation of Interstellar Solar Sail velocities}, Journal of
  the British Interplanetary Society 50~(12) (1997) 475--478.

\bibitem{Zeng2011}
X.~Zeng, J.~Li, H.~Baoyin, S.~Gong, {Trajectory optimization and applications
  using high performance solar sails}, Theoretical and Applied Mechanics
  Letters 1~(3) (2011) 033001.

\bibitem{Dachwald2004}
B.~Dachwald, {Low-thrust trajectory optimization and interplanetary mission
  analysis using evolutionary neurocontrol}, Doktorarbeit, Institut f{\"u}r
  Raumfahrttechnik, Universit{\"a}t der Bundeswehr, M{\"u}nchen.

\bibitem{Dachwald2005}
B.~Dachwald, {Optimal solar sail trajectories for missions to the outer solar
  system}, Journal of Guidance, Control, and Dynamics 28~(6) (2005) 1187--1193.

\bibitem{Kluever1996}
C.~A. Kluever, {Trajectory Optimization of an Interstellar Mission Using Solar
  Electric Propulsion}, NASA~(19980004505).

\bibitem{Abdelkhalik2007}
O.~Abdelkhalik, D.~Mortari, {N-impulse orbit transfer using genetic
  algorithms}, Journal of Spacecraft and Rockets 44~(2) (2007) 456--460.

\bibitem{Heyl2005}
J.~S. Heyl, {The long-term future of space travel}, Physical Review D 72~(10)
  (2005) 107302.
\newblock \href {http://dx.doi.org/10.1103/PhysRevD.72.107302}
  {\path{doi:10.1103/PhysRevD.72.107302}}.

\bibitem{Rindler1960}
W.~Rindler, {Hyperbolic motion in curved space time}, Physical Review 119~(6)
  (1960) 2082--2089.
\newblock \href {http://dx.doi.org/10.1103/PhysRev.119.2082}
  {\path{doi:10.1103/PhysRev.119.2082}}.

\bibitem{Kwan2010}
J.~Kwan, G.~F. Lewis, J.~B. James, {The Adventures of the Rocketeer:
  Accelerated Motion Under the Influence of Expanding Space}, Publications of
  the Astronomical Society of Australia 27~(1) (2010) 15--22.
\newblock \href {http://dx.doi.org/10.1071/AS09050}
  {\path{doi:10.1071/AS09050}}.

\bibitem{Henriques2012}
P.~G. {Henriques}, J.~{Natario}, {The rocket problem in general relativity},
  Journal of Optimization Theory and Applications 154~(2) (2012) 500--524.

\bibitem{Hartle2003}
J.~B. Hartle, {Gravity: An Introduction to Einstein's General Relativity},
  Vol.~1, Addison-Wesley, 2003.

\bibitem{Gasperini2013}
M.~Gasperini, Theory of Gravitational Interactions, Springer, 2013.

\bibitem{Miyamoto1975}
M.~Miyamoto, R.~Nagai, {Three-dimensional models for the distribution of mass
  in galaxies}, Publications of the Astronomical Society of Japan 27 (1975)
  533--543.

\bibitem{Nusser2009}
A.~Nusser, {Ergodic Considerations in the Gravitational Potential of the Milky
  Way}, The Astrophysical Journal 706~(1) (2009) 113--118.
\newblock \href {http://dx.doi.org/10.1088/0004-637X/706/1/113}
  {\path{doi:10.1088/0004-637X/706/1/113}}.

\bibitem{Hernquist1990}
L.~Hernquist, {An analytical model for spherical galaxies and bulges}, The
  Astrophysical Journal 356 (1990) 359--364.

\bibitem{Navarro1997}
J.~F. Navarro, C.~S. Frenk, S.~D. White, {A Universal density profile from
  hierarchical clustering}, The Astrophysical Journal 490~(2) (1997) 493.

\bibitem{Dehnen2006}
W.~Dehnen, D.~E. McLaughlin, J.~Sachania, {The velocity dispersion and mass
  profile of the Milky Way}, Monthly Notices of the Royal Astronomical Society
  369~(4) (2006) 1688--1692.
\newblock \href {http://dx.doi.org/10.1111/j.1365-2966.2006.10404.x}
  {\path{doi:10.1111/j.1365-2966.2006.10404.x}}.

\bibitem{Kafle2014}
P.~R. Kafle, S.~Sharma, G.~F. Lewis, J.~Bland-Hawthorn, {On the Shoulders of
  Giants: Properties of the Stellar Halo and the Milky Way Mass Distribution},
  The Astrophysical Journal 794~(1) (2014) 59.

\bibitem{Bennett2013}
C.~Bennett, D.~Larson, J.~Weiland, N.~Jarosik, G.~Hinshaw, N.~Odegard,
  K.~Smith, R.~Hill, B.~Gold, M.~Halpern, et~al., {Nine-year Wilkinson
  Microwave Anisotropy Probe (WMAP) observations: final maps and results}, The
  Astrophysical Journal Supplement Series 208~(2) (2013) 20.
\newblock \href {http://dx.doi.org/10.1088/0067-0049/208/2/20}
  {\path{doi:10.1088/0067-0049/208/2/20}}.

\bibitem{Bullock2005}
J.~S. Bullock, K.~V. Johnston, {Tracing galaxy formation with stellar halos. I.
  Methods}, The Astrophysical Journal 635~(2) (2005) 931--949.
\newblock \href {http://dx.doi.org/10.1086/497422} {\path{doi:10.1086/497422}}.

\bibitem{Naray2009}
R.~K. de~Naray, S.~S. McGaugh, J.~C. Mihos, {Constraining the NFW potential
  with observations and modeling of low surface brightness galaxy velocity
  fields}, The Astrophysical Journal 692~(2) (2009) 1321.
\newblock \href {http://dx.doi.org/10.1088/0004-637X/692/2/1321}
  {\path{doi:10.1088/0004-637X/692/2/1321}}.

\bibitem{Kafle2012}
P.~R. {Kafle}, S.~{Sharma}, G.~F. {Lewis}, J.~{Bland-Hawthorn}, {Kinematics of
  the Stellar Halo and the Mass Distribution of the Milky Way Using Blue
  Horizontal Branch Stars}, The Astrophysical Journal 761~(1210.7527) (2012)
  98.
\newblock \href {http://dx.doi.org/10.1088/0004-637X/761/2/98}
  {\path{doi:10.1088/0004-637X/761/2/98}}.

\bibitem{Bade1953}
W.~Bade, {Relativistic rocket theory}, American Journal of Physics 21~(4)
  (1953) 310--312.

\bibitem{Betts1998}
J.~T. Betts, {Survey of numerical methods for trajectory optimization}, Journal
  of guidance, control, and dynamics 21~(2) (1998) 193--207.

\bibitem{Vinko2008}
T.~Vink{\'o}, D.~Izzo, {Global optimisation heuristics and test problems for
  preliminary spacecraft trajectory design}, European Space Agency, The
  Advanced Concepts Team.

\bibitem{Charbonneau1995}
P.~Charbonneau, {Genetic algorithms in astronomy and astrophysics}, The
  Astrophysical Journal Supplement Series 101 (1995) 309.
\newblock \href {http://dx.doi.org/10.1086/192242} {\path{doi:10.1086/192242}}.

\bibitem{Coley1999}
D.~A. Coley, {An Introduction to Genetic Algorithms for Scientists and
  Engineers}, World Scientific, 1999.

\bibitem{Kerr1986}
F.~J. Kerr, D.~Lynden-Bell, {Review of galactic constants}, Monthly Notices of
  the Royal Astronomical Society 221~(4) (1986) 1023--1038.

\bibitem{Seeds2015}
M.~A. Seeds, D.~Backman, {Stars and Galaxies}, 9th Edition, Cengage Learning,
  2015.

\bibitem{Moore2011}
P.~Moore, R.~Rees, {Patrick Moore's Data Book of Astronomy}, 2nd Edition,
  Cambridge University Press, 2011.

\bibitem{Jones2004}
M.~H. Jones, R.~J. Lambourne, An Introduction to Galaxies and Cosmology,
  Cambridge University Press, 2004.

\bibitem{Engineering2014}
{Down East Engineering}, Christoffel symbols and geodesics, symbolic model,
  http://au.mathworks.com/matlabcentral/fileexchange/45901-christoffel-symbols-and-geodesics--symbolic-model
  (2014).

\bibitem{Matloff2010}
G.~L. Matloff, {Deep Space Probes: To the Outer Solar System and Beyond}, 2nd
  Edition, Springer, 2010.

\bibitem{Westmoreland2010}
S.~Westmoreland, {A note on relativistic rocketry}, Acta Astronautica 67~(9)
  (2010) 1248--1251.
\newblock \href {http://dx.doi.org/10.1016/j.actaastro.2010.06.050}
  {\path{doi:10.1016/j.actaastro.2010.06.050}}.

\bibitem{Tinder2006}
R.~F. Tinder, {Relativistic Flight Mechanics and Space Travel: A Primer for
  Students, Engineers and Scientists (Synthesis Lectures on Engineering
  Series)}, Morgan and Claypool Publishers, 2006.

\bibitem{Baumgardt2006}
H.~Baumgardt, A.~Gualandris, S.~P. Zwart, {Ejection of hypervelocity stars from
  the Galactic Centre by intermediate-mass black holes}, Monthly Notices of the
  Royal Astronomical Society 372~(1) (2006) 174--182.

\bibitem{Hills1988}
J.~G. Hills, {Hyper-velocity and tidal stars from binaries disrupted by a
  massive Galactic black hole}, Nature Publishing Group.

\end{thebibliography}

\end{document}